\documentclass[aps,prd,preprintnumbers,nofootinbib,twocolumn,10pt]{revtex4-2}

\usepackage{amsfonts,amsmath,amssymb}
\usepackage{graphicx,epsfig}
\usepackage{float}
\usepackage{subfig}

\def	\be	{\begin{equation}}
\def	\ee	{\end{equation}}

\def	\eps	{\epsilon}

\newcommand*{\affmark}[1][*]{\textsuperscript{#1}}

\begin{document}

\title{A Local First Law of Gravity}

\author{Maulik Parikh\affmark[1], Sudipta Sarkar\affmark[2], and Andrew Svesko\affmark[1]}

\affiliation{\affmark[1]Department of Physics and Beyond: Center for Fundamental Concepts in Science\\
Arizona State University, Tempe, Arizona 85287, USA\\
\affmark[2]Indian Institute of Technology, Gandhinagar 382355, Gujarat, India}

\begin{abstract}
\begin{center}
{\bf Abstract}
\end{center}
\noindent
We find a (quasi-)local first law of thermodynamics, $\Delta E = T \Delta S - W$, connecting gravitational entropy, $S$, with matter energy and work. For Einstein gravity $S$ is the Bekenstein-Hawking entropy, while for general theories of gravity $S$ is the Wald entropy, evaluated on the stretched future light cone of any point in an arbitrary spacetime, not necessarily containing a black hole. The equation can  be written as $\rho \Delta V = T \Delta S - p \Delta V$ by regarding the energy-momentum tensor as that of a fluid.
\end{abstract}

\thispagestyle{empty}

\maketitle

\noindent
\emph{Introduction.} \textendash{}
For macroscopic systems of matter, the first law of thermodynamics 
states that
\be 
\Delta E = T \Delta S_{\rm rev} - W \; , \label{firstlawthermo}
\ee
where, by the Clausius theorem, $\Delta S_{\rm rev} = Q/T$ is the reversible component of the change in entropy. A similar equation holds for a Schwarzschild black hole \cite{bardeen}:
\be
\Delta M = T \Delta S \; . \label{bhfirstlaw}
\ee
Here $M$ is the Arnowitt-Deser-Misner (ADM) mass of the black hole, $T$ is its Hawking temperature, and the Bekenstein-Hawking entropy, $S$, is a quarter of the area of the event horizon, measured in Planck units \cite{Bekenstein:1973ur}:
\be
S=\frac{A}{4G \hbar} \; . \label{SBH}
\ee
Equation (\ref{bhfirstlaw}) is the famous first law of black hole thermodynamics.

Despite the superficial similarities between (\ref{firstlawthermo}) and (\ref{bhfirstlaw}), these expressions are rather different in character. First of all, the black hole law only applies, obviously, in the presence of a black hole. Also, unlike (\ref{firstlawthermo}), the black hole law is not local: the definition of an event horizon in general relativity involves the global causal structure of spacetime. Moreover, a formal definition of the mass term calls for special asymptotic boundary conditions, in particular asymptotic flatness; generically, energy density cannot simply be integrated over finite regions of space to obtain the total energy. Hence the left-hand side of (\ref{bhfirstlaw}) has no exact definition for the realistic case of, say, an astrophysical, uncharged black hole in an expanding universe. Another distinction is that, whereas in equation (\ref{firstlawthermo}) the system can exchange energy with a thermal reservoir, there is no physical process \cite{Gao:2001ut,Jacobson:2003wv} by which the ADM mass can change because the total energy at spacelike infinity in an asymptotically flat spacetime is a conserved quantity. Instead, the $\Delta M$ in (\ref{bhfirstlaw}) refers to differences in the ADM mass under a variation in the space of static uncharged black hole solutions. Finally, the work term is notably absent in (\ref{bhfirstlaw}); indeed, neither pressure nor spatial volume admits a straightforward definition for black holes \cite{Parikh:2005qs,Dolan:2011xt,Dolan:2012jh,Kubiznak:2016qmn}.

The aim of this paper is to derive a local first law of thermodynamics that also includes gravitational entropy. The main result is the discovery of a hybrid equation,
\be
\Delta E=T \, \Delta \! \left ( \frac{A_{\rm rev}}{4G \hbar} \right )- W \; , \label{localfirst}
\ee
combining attributes of (\ref{firstlawthermo}) and (\ref{SBH}). We find that such an equation applies, within a suitably defined region, to all matter-gravity systems that are significantly smaller than the local curvature scale of spacetime. To arrive at (\ref{localfirst}), we shall incorporate three uncommon elements. First, energy will be measured with respect to accelerating observers, rather than with respect to inertial observers. Second, we will consider a two-sphere of accelerating observers with constant uniform radially-outward acceleration \cite{Parikh:2017}. These observers collectively sweep out a hypersurface that asymptotes to the future light cone of the point at the center of the two-sphere, generating a kind of stretched future light cone (Figure 1). Third, we will use Einstein's equation to convert the heat flux through the hypersurface into the change in gravitational entropy. But since outward-accelerating observers will spread out even in the absence of heat flux (such as in Minkowski space), we will find that only the reversible part of the change in gravitational entropy is associated with the heat flux. The corresponding area change of the hypersurface is denoted $\Delta A_{\rm rev}$ in (\ref{localfirst}). This quite-general construction can be set up about an arbitrary point in an arbitrary spacetime; our derivation uses no speculative assumptions beyond the validity of quantum field theory in curved spacetime. Amusingly (and somewhat mysteriously), we can express (\ref{localfirst}) in terms of fluid properties as
\be
\rho \Delta V = T \, \Delta \! \left ( \frac{A_{\rm rev}}{4G \hbar} \right ) - p \Delta V \; , \label{fluidlaw}
\ee
where $\rho$ and $p$ are the energy density and pressure measured by inertial observers, and $V$ is the volume of a ball in Euclidean space, namely $\frac{4}{3} \pi r^3$. Lastly, we also find that an analogous expression holds for more general theories of gravity, in which the Bekenstein-Hawking entropy is replaced by the Wald entropy \cite{Wald:1993nt}.

\noindent
\emph{Geometric Set-up.} \textendash{} 
In the neighborhood of any spacetime point, $P$, the metric can always be expanded around that of flat space. In Riemann normal coordinates, 
\be
g_{ab}(x) = \eta_{ab} - \frac{1}{3} R_{acbd}(P) x^c x^d + \dots \; ,\label{RNC}
\ee
where $x^a$ are Cartesian coordinates with their origin at $P$. Using the local Lorentz symmetry, pick an arbitrary time, $t$, to split spacetime into space and time. The local Poincar\'e symmetries are generated by approximate Killing vectors, which for a generic spacetime fail to exactly obey the Killing equation because of the quadratic terms in (\ref{RNC}). For example, the vector $x \partial^a_t + t \partial^a_x$ generates a Cartesian boost in the $x$-direction, and is an approximate Killing vector \cite{Jacobson:1995ab}. We will instead use a vector field that generates radial boosts:
\be
\xi^{a}=r\partial^{a}_{t}+t\partial^{a}_{r} \; . \label{KV}
\ee
Of course, radial boosts are not isometries even in Minkowski space \cite{DeLorenzo:2017tgx}, but notice that $\xi^a$ does, to leading order in Riemann normal coordinates, satisfy some -- though not all -- components of Killing's equation:
\be
\begin{split}
&\nabla_{t}\xi_{t}=0 + {\cal O}(x^2)\;,\quad \nabla_{t}\xi_{i}+\nabla_{i}\xi_{t}=0 + {\cal O}(x^2)\;,\\
&\nabla_{i}\xi_{j}+\nabla_{j}\xi_{i}=\frac{2t}{r}\left(\delta_{ij}-\frac{x_{i}x_{j}}{r^{2}}\right) + {\cal O}(x^2) \; ,	\label{Killingfailure}
\end{split}
\ee
where the $x^i$ are Cartesian spatial Riemann normal coordinates and we have $r = \sqrt{x^i x_i}$ and $\partial_r^a = \frac{x^i}{r} \partial^a_i$. 

In the absence of spacetime curvature, $\xi^a$ is tangent to the flow lines of observers with constant proper acceleration in the outward radial direction. The congruence of such worldlines with uniform acceleration $1/\alpha$ forms a hyperboloid asymptoting to the light cone at $P$. The hyperboloid is described by $r^2 = \alpha^2 + t^2$; its constant-time sections are two-spheres of radius $r(t)$ bounding a three-ball $B(t)$ (Figure 1). On the hyperboloid, $\xi^2 = -\alpha^2$.

\begin{figure}[H]
\includegraphics[width=8.5cm]{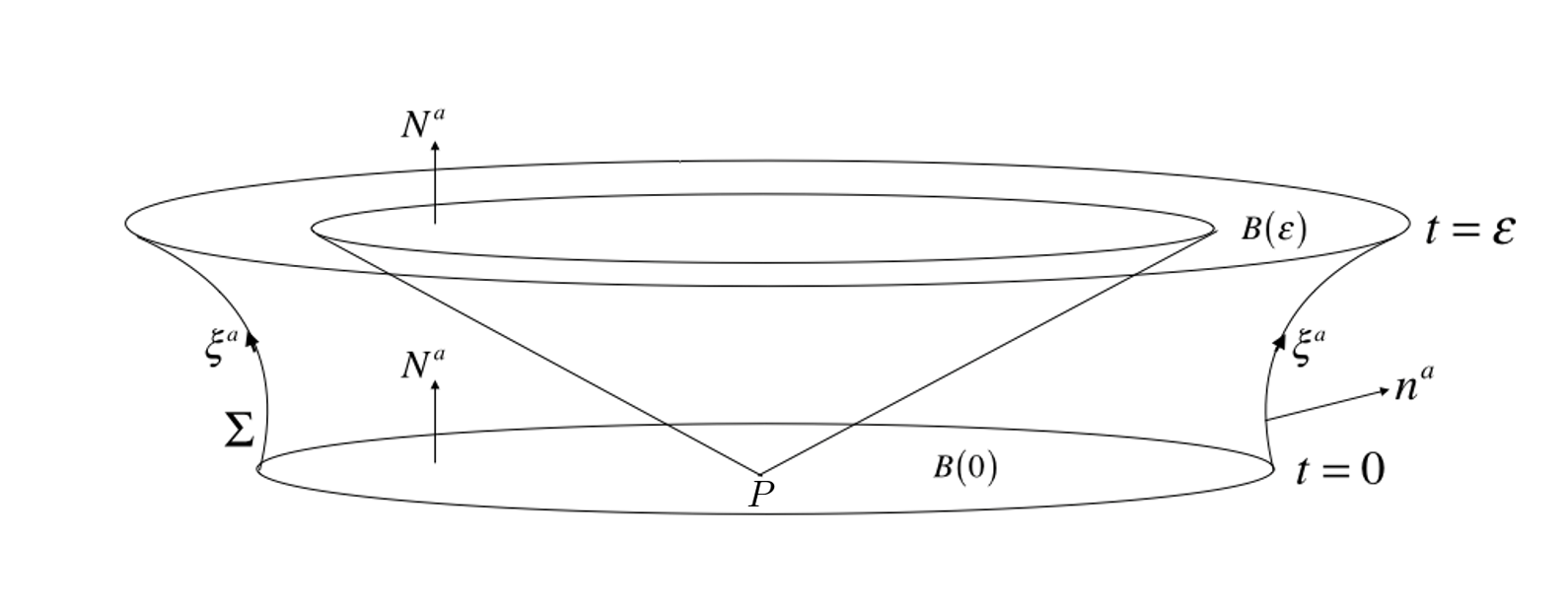}
\centering
\caption{Consider an arbitrary point $P$ in spacetime and choose a locally inertial time coordinate, $t$. Radially accelerating worldlines $\xi^{a}$ with uniform acceleration generate a timelike congruence, $\Sigma$, with unit outward-pointing normal $n^a$. Two constant-time sections bounded by $\Sigma$ at time $t = 0$ and $t = \epsilon$ are indicated by the three-balls $B(0)$ and $B(\epsilon)$, and have unit normal $N^a$, which is taken to be future-directed. The boundary of the four-volume, $M$, containing our thermodynamic system is $\Sigma \cup B(0) \cup B(\epsilon)$.}
\label{hyperboloid}
\end{figure}

The presence of spacetime curvature necessitates a more careful treatment \cite{Parikh:2017}. We replace the hyperboloid by a timelike surface $\Sigma$, defined as follows. Let $\alpha$ be a small scale (small compared with the smallest curvature scale at $P$) with dimensions of length. Imagine that the radial boost vector field $\xi^a$, defined in (\ref{KV}), describes the unnormalized tangent vectors to the worldlines of a set of observers. Select the subset of observers who have instantaneous proper acceleration $1/\alpha$ at time $t = 0$; this subset forms a small deformation of the coordinate two-sphere $r = \alpha$.
Next, choose another dimensionful scale $\epsilon$ and follow the chosen observers from $t = 0$ to $t = \epsilon$. In general, because of spacetime curvature, the observers will not maintain their acceleration. However, by evolving for a very short period of time, $\epsilon \ll \alpha$, the acceleration can be treated as effectively constant. We can therefore, in the presence of spacetime curvature, regard $\Sigma$ as the worldtube of nearly-uniformly accelerated observers (Figure 1). The normal to $\Sigma$ is
\be
n^a = \frac{t}{\alpha} \partial_t^a + \frac{r}{\alpha} \partial_r^a + \dots \; .
\ee
In general, the effect of spacetime curvature is to make $\Sigma$ a small deformation of the hyperboloid $r^2 = \alpha^2 + t^2$, and to limit its extent in time, $0 \leq t \leq \epsilon \ll \alpha$. The dominant contribution to most calculations will come from taking $\Sigma$ to be the hyperboloid; only for the entropy change will deviations from the hyperboloid be implicitly important. Our thermodynamic system consists of the matter and gravity contained within the four-volume, $M$, bounded by $\Sigma$ and the constant-time planes $t = 0$ and $t = \epsilon$. 

The motivation for this set-up is two-fold \cite{Parikh:2017}. The compactness of cross-sections of $\Sigma$ means that three-balls at any given time can have finite volume and surface area. Also, by the Davies-Unruh effect, proper acceleration is related to temperature:
\be
T = \frac{\hbar}{2 \pi \alpha} \; . \label{temp}
\ee
Hence $\Sigma$ has a uniform temperature, $T$. This follows from the choice of the Poincar\'e-invariant state, which is motivated by the strong principle of equivalence: free-falling observers should see the same physics locally as inertial observers in Minkowski space. But any coherent state other than the Poincar\'e-invariant vacuum would lead to a stress tensor whose vacuum expectation value would be singular somewhere. Given the Poincar\'e-invariant state, the thermality of the surface $\Sigma$, which in spherical Rindler coordinates is approximately a constant-lapse surface with lapse $\alpha$, follows geometrically (at least formally) from the local Tolman temperature given by Eq. (\ref{temp}). 

It is rather more subtle to give the temperature an operational meaning in terms of Rindler observers and model Unruh detectors. For such detectors, it is known \cite{Barbado:2012fy,Doukas:2013noa,Fewster:2015dqb} that to actually observe a perfectly thermal spectrum requires an eternally accelerating trajectory. An analysis \cite{Barbado:2012fy} of transient acceleration shows that a thermal spectrum is detected to arbitrary accuracy provided the duration of acceleration is sufficiently long compared with the inverse acceleration. This condition is possible to arrange, albeit at a price. We extend the worldlines of the observers over a longer time, $\tau$, much greater than the inverse acceleration, $\alpha$ (but still short enough that curvature effects are negligible), keeping the proper acceleration constant. As $\alpha$ can be arbitrarily small, this is always possible. Our surface $\Sigma$ is then a brief segment, $0 < t < \epsilon \ll \alpha \ll \tau$ of an extended surface traced by a congruence of such observers. (In general, the worldlines will not trace the integral curves of  $\xi^a$ before $t = 0$ or after $t = \epsilon$ and we therefore restrict our calculation to $\Sigma$.) The class of observers who continue to accelerate at $1/\alpha$ on the extended surface beyond $\Sigma$ will eventually register a roughly thermal spectrum, with a temperature (\ref{temp}) matching their proper acceleration. 

This construction shows that certain observers will indeed detect a thermal spectrum corresponding to a geometric concept, namely the surface gravity of $\Sigma$, as already determined by the vacuum state. However, the price of giving an operational meaning to the temperature is that the geometric construction is no longer local, but only quasi-local. We leave to future work further clarification of the physical interpretation of the proper acceleration along $\Sigma$ as temperature.

Our construction is reminiscent of a similar formulation based on spherical Rindler coordinates that was proposed in \cite{Kothawala:2014oba} to define an equi-geodesic hyperboloid; that surface, however, does not generically coincide with our isothermal surface, $\Sigma$.\\

\noindent
\emph{A First Law for Matter.} \textendash{} 
The radially-accelerating observers have a normalized four-velocity vector $u^a \equiv \xi^a/(-\xi^2)^{1/2} \approx \xi^a/\alpha$, to leading order. Let the energy-momentum tensor be $T_{ab}$. Then the energy current measured by the accelerating observers is
\be
J^a =-T^{ab}u_{b} = -\frac{1}{\alpha} T^{ab} \xi_b \; . \label{energycurrent}
\ee
If $\xi^a$ were a Killing vector, this current would be conserved by Killing's equation. However, since $\xi^a$ is not a Killing vector, we have
\be 
\int_{M}d^{4}x\nabla_{a}J^{a} = -\frac{1}{\alpha}  \int_{M}d^{4}x T^{ab} \nabla_a \xi_b \; .
\ee
Applying the divergence theorem to the left-hand side and rearranging, we find
\be
\begin{split}
& \frac{1}{\alpha}\int_{B(\epsilon)} dS_a T^{ab}\xi_{b} 
- \frac{1}{\alpha}\int_{B(0)} dS_a T^{ab}\xi_{b}
\\
& \qquad = \frac{1}{\alpha}\int_{\Sigma}d\Sigma_{a}T^{ab}\xi_{b} 
 - \frac{1}{\alpha}\int_{M}d^{4}xT^{ab}\nabla_{a}\xi_{b} \; , \label{stokes}
\end{split} 
\ee
where, in accordance with Stokes' theorem, the signs depend on whether a boundary is timelike or spacelike. Here $dS^{a}=N^{a}d^3x = \partial_t^a r^2 dr d \Omega$ and $d\Sigma^{a}=n^{a}d^3x \approx n^a dt (\alpha/r) r^2(t) d \Omega$, where $dt (\alpha/r)$ is the differential of proper time on the hyperboloid. We now argue that these terms can be interpreted as the change in energy, the heat flow, and the work done, so that (\ref{stokes}) is the first law of thermodynamics for matter. 

It is evident that $E(t)$, the energy of the system at time $t$, is given by $\frac{1}{\alpha}\int_{B(t)}dS_a T^{ab}\xi_{b}$, where $B(t)$ is the three-ball section of ${\cal M}$ at constant $t$. Not only does this expression have the correct dimension of energy, but $E(t)$ is simply the Noether charge associated with the energy current density, (\ref{energycurrent}). We then find that the difference between the energy at $t = \epsilon$ and $t = 0$ is
\be 
\Delta E = \frac{1}{\alpha}\int_{B(\epsilon)}dS_a T^{ab}\xi_{b}  - \frac{1}{\alpha}\int_{B(0)}dS_a T^{ab}\xi_{b} \; ,     \label{energy}
\ee
which is indeed the expression on the left-hand side of (\ref{stokes}). It is interesting to evaluate $\Delta E$ explicitly. We first note that, to leading order in Riemann normal coordinates, the energy-momentum tensor $T^{ab}(x) = T^{ab}(P) + {\cal O}(x)$ can be replaced within the integral by its value at $P$. Referring to (\ref{KV}), we then see that the off-diagonal pieces of $T^{ab}$ integrate to zero because the integral of a Cartesian spatial coordinate over a ball centered at the origin vanishes. We are therefore left with $E(t) = \frac{4 \pi}{\alpha} T^{tt}(P) \int_0^{r(t)} dr r^2 N_t \xi_t$. We can approximate the radius of the ball by the radius of the hyperboloid. Hence $\Delta E = 2 \pi T^{tt}(P) \alpha \epsilon^2$, using also $\epsilon \ll \alpha$. Similarly, the volume of $B(t)$ is $V(t) = \frac{4}{3} \pi (\alpha^2 + t^2)^{3/2}$. Then the difference between the volume of $B(\epsilon)$ and of $B(0)$ is
\be
\Delta V = 2 \pi \alpha \eps^2 \; . \label{volume}
\ee
Labelling the energy density $\rho \equiv T^{tt}(P)$, we obtain 
\be
\Delta E = \rho \Delta V \; . \label{rhodV}
\ee
It is amusing that, even though $\Delta E$ is the difference in energies as measured by accelerating observers, it can nevertheless be written in terms of $\rho$ and $\Delta V$, the energy density and volume change measured by inertial observers; it is not the case, though, that $E(t) = \rho V(t)$.

Next, consider the first term on the right in (\ref{stokes}). This is clearly the integrated energy flux into the timelike surface $\Sigma$. The sign matches too: the normal to $\Sigma$ is outward-pointing, while the energy current, $J^a$, is defined with a minus sign, (\ref{energycurrent}). Now, in thermodynamics, heat is the energy flowing into macroscopically unobservable degrees of freedom. For our observers on the stretched future light cone, the interior of the system is fundamentally unobservable, being causally disconnected. We can therefore interpret the integrated energy flux into the system as heat \cite{Jacobson:1995ab}:
\be 
Q=\frac{1}{\alpha}\int d\Sigma_{a}T^{ab}\xi_{b} \; .   \label{deltaQ}
\ee
This interpretation will be confirmed when we incorporate gravity. 

Finally, consider the last term in (\ref{stokes}). At first sight, this term does not appear to be a work term because it is an integral over a four-volume. To see that it is, consider first for simplicity a diagonal energy-momentum tensor with isotropic pressure, $T^{ij}(P)=p\delta^{ij}$. Then, working as always at leading order, we find
\be 
\frac{1}{\alpha}\int_{M}d^{4}xT^{ab}\nabla_{a}\xi_{b}\approx\frac{1}{\alpha}\int_{M}d^{4}x\frac{2pt}{r} \approx 2\pi p\alpha\epsilon^{2} \; ,
\ee
where, in the last step, we have evaluated the integral at leading order in $\epsilon$. From (\ref{volume}), we see that this is exactly equal to $p \Delta V$, the pressure-volume work done by a system, motivating the identification of the last term in (\ref{stokes}) as work. For previous proposals for a $p \Delta V$ work term see \cite{Padmanabhan:2002sha,Chakraborty:2015aja}. However, in those works, neither the volume nor the pressure match our definitions.\\

More generally, consider an arbitrary energy-momentum tensor, for which $T^{ii}(P)=p_{i}$, and $T^{ij}\neq0$ for $i\neq j$. Now from (\ref{Killingfailure}), we have $\partial_i \xi_j \sim \frac{t}{r^{3}}x_i x_j$ for $i \neq j$. This is an odd function of the coordinates and therefore $T^{ij} \partial_i \xi_j$ vanishes under integration over the three-ball for $i \neq j$. 
Moreover, $T^{xx}\partial_{x}\xi_{x}=p_{x}\frac{t}{r^{3}}\left(y^{2}+z^{2}\right)$, and similarly for $T^{yy}$ and $T^{zz}$. Then we find
\be
W = \frac{1}{\alpha}\int_{M}d^{4}xT^{ab}\nabla_{a}\xi_{b} = \left (\frac{1}{3} \sum_{i=1}^{3}p_{i} \right ) \Delta V \; , \label{work}
\ee
which is precisely the pressure-volume work for anisotropic pressures, and is now valid for arbitrary energy-momentum tensors. 

Consulting (\ref{energy}), (\ref{deltaQ}), and (\ref{work}), we indeed find that (\ref{stokes}) can be interpreted as a first law of thermodynamics for accelerating observers moving along $\Sigma$. Our first law is local in that it is valid near an arbitrary point in a generic spacetime. As it stands though, this equation does not yet involve gravity: there is no Newton's constant and all the terms involve the energy-momentum tensor of matter, $T^{ab}$. To turn it into a local first law with gravity, we now invoke Einstein's equation.

\noindent
\emph{Connecting Matter and Spacetime Thermodynamics.} \textendash{}  
Using Einstein's equation, $R_{ab} - \frac{1}{2} R g_{ab} + \Lambda g_{ab} = 8 \pi G T_{ab}$, in (\ref{deltaQ}) we find $Q = \frac{1}{8 \pi G \alpha}\int_{\Sigma} d\Sigma_{a} R^a_e \xi^{e}$. The terms proportional to the metric vanish when contracted with $d \Sigma_a$ and $\xi_b$ because $\xi^a$ lies along $\Sigma$ while $n^a$ is normal to it.

Now if $\xi^a$ were a Killing vector, it would obey Killing's identity: $\nabla_b \nabla_c \xi_d = R_{ebcd} \xi^e$. However, we already know that $\xi^a$ is not exactly a Killing vector. We therefore have $\nabla_b \nabla_c \xi_d - R_{ebcd} \xi^e = f_{bcd}$ where $f_{bcd}$ encodes the failure of Killing's identity to hold. Then
\be
Q =\frac{1}{8\pi G \alpha}\int_{\Sigma}d\Sigma_{a} \frac{1}{2} (g^{ac} g^{bd} - g^{ad}g^{bc})  ( \nabla_{b}\nabla_{c}\xi_{d} - f_{bcd} ) \; . \label{Qint}
\ee 
We now show that the integral of the $\nabla_{b}\nabla_{c}\xi_{d}$ term evaluates to $T \Delta S$, by essentially reversing the thermodynamic derivation of Einstein's equations in the Noether charge approach \cite{Parikh:2009qs,Guedens:2011dy,Parikh:2017}. First, we use Stokes' theorem for an antisymmetric tensor field $A^{ab}$,
namely $\int_{\Sigma}d\Sigma_{a}\nabla_{b}A^{ab}=-\oint_{\partial\Sigma}dS_{ab}A^{ab}$, to express that integral as the difference of terms $-\frac{1}{8 \pi G \alpha} \int dS_{ab} \frac{1}{2} (g^{ac} g^{bd} - g^{ad}g^{bc}) \nabla_{c}\xi_{d}$ evaluated over the two-spheres at time $t = 0$ and $t=\epsilon$. Here $dS_{ab} = dA \frac{1}{2} (n_a u_b - u_a n_b)$. Then, since $u^a \approx \xi^a/\alpha$, we have
\be
-\frac{1}{16 \pi G \alpha^2} \int dA (n^c \xi^d - n^d \xi^c) \nabla_c \xi_d = +\frac{A}{8 \pi G \alpha} = T \frac{A}{4 G \hbar} \; .
\ee
Here we used the fact, (\ref{Killingfailure}), that the projection of $\nabla_c \xi_d$ in the $n-\xi$ plane is antisymmetric. We then made use of our judicious choice of $\Sigma$ as a surface of constant acceleration and thus temperature in writing $\xi^c \nabla_c \xi^d = \alpha n^d$ and in using (\ref{temp}). Hence the integral of the $\nabla_{b}\nabla_{c}\xi_{d}$ term can be written as $T \Delta S$, where $S$ is precisely the Bekenstein-Hawking entropy, suggesting that gravitational entropy can be associated with sections of $\Sigma$.

Now consider the $f_{bcd}$ term in the $Q$ integral, (\ref{Qint}). In general, $f_{bcd}$ consists of two types of terms: terms that arise because of spacetime curvature, (\ref{RNC}), but also a term of ${\cal O}(x^{-1})$ that exists even in Minkowski space, because radial boosts are not true isometries; explicitly, the latter comes from taking partial derivatives of (\ref{Killingfailure}). To lowest order, the integrals of terms of the first type either vanish because they are linear Cartesian spatial terms integrated over a sphere \cite{Parikh:2017} or, if not, can be canceled by adding quadratic and cubic terms to $\xi^a$ \cite{Guedens:2012sz,Guedens:2011dy,Parikh:2017}.

The integral of the term of ${\cal O}(x^{-1})$ in $f_{bcd}$ cannot be eliminated by re-definitions of $\xi^a$. To leading order, we have $f_{bcd}=\partial_{b}\partial_{c}\xi_{d}$. Then, direct calculation yields $\frac{1}{8\pi G \alpha}\int_{\Sigma}d\Sigma_{a} \frac{1}{2} (\eta^{ac} \eta^{bd} - \eta^{ad}\eta^{bc})  (- f_{bcd} ) = - \frac{\epsilon^2}{2 G \alpha}$.
Hence
\be
Q = T \Delta S - \frac{\epsilon^2}{2 G \alpha} \; . \label{delQrev}
\ee
To understand the second term, consider the change in entropy of the hyperboloid defined by $r^2 = \alpha^2 + t^2$:
\be
T\Delta S_{\rm hyp}
=\frac{\hbar}{2 \pi \alpha} \frac{1}{4 G \hbar} (A_{\rm hyp}(\epsilon)-A_{\rm hyp}(0) )
= \frac{\epsilon^2}{2 G \alpha} \; , \label{TDeltaSMink}
\ee
where $A_{\rm hyp}(t) = 4 \pi (\alpha^2 + t^2)$ is simply the area of constant-$t$ sections of the hyperboloid. We see that (\ref{delQrev}) automatically subtracts off the entropy increase from the background expansion of the hyperboloid. Now, in thermodynamics, $Q$ is equal to $T\Delta S$ only for reversible processes. The expansion of the hyperboloid is, like the free expansion of a gas or of the light cone, an irreversible process unrelated to the presence of any heat flux. Hence
\be 
Q = T \Delta S- T\Delta S_{\rm hyp} \equiv T \Delta S_{\rm rev} \; .
\ee
Here $\Delta S_{\rm rev}$ is the reversible part of the change in gravitational entropy. A direct calculation using (\ref{deltaQ}) shows that $Q = (\rho + \frac{1}{3}\sum_i p_i) \Delta V$. Hence we have that $\Delta S_{\rm rev} \geq 0$ if the null energy condition holds.

Putting everything together, we arrive at our result:
\be
\Delta E=T \, \Delta \! \left ( \frac{A_{\rm rev}}{4G \hbar} \right )- W \; . \label{localfirstlaw}
\ee
We have found a hybrid first law that resembles both the ordinary first law of thermodynamics for matter (in that it is valid locally and has a work term) as well as the first law for black holes (in that it involves gravitational entropy). Using (\ref{rhodV}) and (\ref{work}), we can also put this in the form (\ref{fluidlaw}). In (\ref{localfirstlaw}), $\Delta E$ and $W$ refer to the energy of and work done by matter, while the middle term refers to the entropy of gravity. The result suggests that (stretched) future light cones possess thermodynamic entropy, which is perhaps not unreasonable as their interiors are causally disconnected from the outside. Note the absence of a term corresponding to the entropy of matter. This property is reminiscent of black holes: if one empties a cup of hot coffee into a black hole, the black hole's entropy increases solely due to the mass-energy of the coffee, with no extra contribution from the coffee's own thermal entropy. It is also notable that, because all terms vanish when $T_{ab}$ is zero, there is no contribution of gravitational energy in our local first law; indeed, inclusion of such energy would require a quasi-local conservation law \cite{McGrath:2012db,Epp:2013hua}. 
\\
 
Our result also bears some resemblance to recent work \cite{Jacobson:2015hqa,Bueno:2016gnv}, in which the assumption that the vacuum is maximally entangled leads to the ``first law of causal diamond mechanics," including a term that can be interpreted as work. It would be interesting to explore this connection, as well as to some of the ideas expressed in \cite{Padmanabhan:2009vy}.
\\

The local first law can be extended to higher-dimensional spacetime; in particular, (\ref{delQrev}) always corresponds to subtracting the inherent area increase of the hyperboloid. More significantly, the derivation can also be extended to a broad class of higher-curvature theories of gravity. Consider a diffeomorphism-invariant gravitational theory for which the Lagrangian, $L$, is a polynomial in the Riemann tensor. Define $P^{abcd}=\partial L/\partial R_{abcd}$. The generalization of Einstein's equation is then $P_a^{cde} R_{bcde} - 2 \nabla^c \nabla^d P_{acdb} - \frac{1}{2} L g_{ab} = 8 \pi G T_{ab}$. Substituting this into $Q$, and replacing $R_{ebcd} \xi^e$ as before with $\nabla_b \nabla_c \xi_d - f_{bcd}$, we again find that the $f_{bcd}$ integral precisely cancels the background increase in gravitational entropy of the hyperboloid. The other term integrates via Stokes' theorem to give the difference in a formally Wald-like gravitational entropy,
\be 
S^{\rm Wald}=-\frac{1}{4G \hbar}\int dS_{ab}(P^{abcd}\nabla_{c}\xi_{d}-2\xi_{d}\nabla_{c}P^{abcd})  \; ,   \label{Waldent}
\ee
which would be exactly the Wald entropy \cite{Wald:1993nt} if $\xi^a$ were a timelike Killing vector in a black hole spacetime. Hence we again find a local first law that takes precisely the form $\Delta E = T \Delta S^{\rm Wald}_{\rm rev} - W$.

Historically, the laws of black hole mechanics supported, as an analogy, Bekenstein's idea that a black hole could be attributed thermodynamic entropy proportional to the horizon area; this was found to be literally true with the discovery that black holes have temperature. Here we have shown that the first law holds locally when restricted to brief segments of stretched future light cones generated by families of accelerating observers. Each term in our first law has independent justification. This result supports an analogy between (quasi-)locally-defined geometric properties and thermodynamic quantities, notably entropy. However, since it is already known that accelerating observers perceive a temperature, our result suggests that stretched future light cones can be regarded literally as having thermodynamic entropy. It would be interesting to know whether this has a statistical-mechanical origin.

\bigskip
\noindent
{\bf Acknowledgments} \\
\noindent
We would like to thank Ted Jacobson, David Kubiznak, Stefano Liberati, and Sergey Solodukhin for helpful comments on the manuscript. MP is supported in part by John Templeton Foundation grant 60253 and by the Government of India DST VAJRA Faculty Scheme VJR/2017/000117. SS is supported in part by the Government of India DST under Start-up grant for Young Scientists YSS/2015/001346.

\end{document}